\documentclass[usenatbib]{aa}
\usepackage{txfonts}
\usepackage{graphicx}
\usepackage[update,prepend]{epstopdf}

\begin{document}

\title{Prospects for detecting the astrometric signature of~Barnard’s~Star~b}

\author{
L.\,Tal-Or\inst{1,5}
\and S.\,Zucker\inst{1}
\and I.\,Ribas\inst{2,3}
\and G.\,Anglada-Escud\'e\inst{4}
\and A.\,Reiners\inst{5}
}

\institute{Department of Geophysics, Raymond and Beverly Sackler Faculty of Exact Sciences, Tel Aviv University, Tel Aviv, 6997801, Israel
\email{levtalo@tauex.tau.ac.il}
\and Institut de Ci\`encies de l’Espai (ICE, CSIC), Campus UAB, c/ de Can Magrans s/n, E-08193 Bellaterra, Barcelona, Spain
\and Institut d’Estudis Espacials de Catalunya (IEEC), E-08034 Barcelona, Spain
\and School of Physics and Astronomy, Queen Mary, University of London, 327 Mile End Road, London, E1 4NS
\and Institut f{\"u}r Astrophysik, Georg-August Universit{\"a}t, Friedrich-Hund-Platz 1, 37077 G{\"o}ttingen, Germany
}

\date{Received 14 November 2018 ; Accepted 09 January 2019}
\abstract
{A low-amplitude periodic signal in the radial velocity (RV) time series of Barnard's Star was recently attributed to a planetary companion with a minimum mass of ${\sim}3.2$\,M$_\oplus$ at an orbital period of ${\sim}233$ days. The relatively long orbital period and the proximity of Barnard's Star to the Sun raises the question whether the true mass of the planet can be constrained by accurate astrometric measurements. By combining the assumption of an isotropic probability distribution of the orbital orientation with the RV analysis results, we calculated the probability density function of the astrometric signature of the planet. In addition, we reviewed the astrometric capabilities and limitations of current and upcoming astrometric instruments. We conclude that Gaia and the Hubble Space Telescope (HST) are currently the best-suited instruments to perform the astrometric follow-up observations. Taking the optimistic estimate of their single-epoch accuracy to be $\sim30$\,$\mu$as, we find a probability of $\sim10\%$ to detect the astrometric signature of Barnard's Star b with $\sim50$ individual-epoch observations. In case of no detection, the implied mass upper limit would be ${\sim}8$\,M$_\oplus$, which would place the planet in the super-Earth mass range. In the next decade, observations with the Wide-Field Infrared Space Telescope (WFIRST) may increase the prospects of measuring the true mass of the planet to ${\sim}99\%$.
}
\authorrunning{Tal-Or et al.}
\titlerunning{Prospects for detecting the astrometric signature of Barnard’s Star b}
\maketitle

\section{Introduction}
\label{sec1}

Nearly $4,000$ exoplanets have been discovered in the last three decades\footnote{http://exoplanet.eu/catalog/}. Transiting-planet surveys, such as the NASA {\it Kepler} mission \citep{Borucki2008}, have provided ${\sim}75\%$ of these discoveries, while the radial velocity (RV) technique has been used to discover another ${\sim}20$\% of them. These discoveries have enabled the estimation of the planet occurrence rates around FGKM dwarf stars \citep[e.g.,][]{Howard2010Sci,Howard2012,DressingCharbonneau2013}. It was found that ${{\sim}50\%}$ of FGK dwarfs, and virtually all M dwarfs, harbor small planets ($1$--$4$\,R$_{\oplus}$) in orbital periods $\lesssim 1$ year \citep[e.g.,][]{WinnFabrycky2015}.

Focusing on the ${\sim}50$ stellar systems within $5$\,pc from the Sun, we find that planets were discovered in $\lesssim20$ of these system. If nearby stars follow the above occurrence rates, there are ${\sim}30$ Earth-to-Neptune size planets in orbital periods $\lesssim 1$\,year yet to be discovered. Assuming an isotropic distribution of orbit orientations, we do not expect to find more than one of these planets via transit search. The most promising way to find the yet-undetected planets would currently be the RV approach. The expected RV semi-amplitudes are on the order of $1$\,m\,s$^{-1}$, which is at the detection limit of the currently available RV instruments such as HARPS \citep{mayor03} and CARMENES \citep{Quirrenbach2018}. However, detecting the RV signals of these planets is complicated by the fact that most of the nearby stars are intrinsically faint M dwarfs and many are also magnetically active \citep[e.g.,][]{SuarezMascareno2017}.

Planets found around nearby stars are valuable targets for follow-up studies. The proximity of these planetary systems to the Sun makes many of their planets accessible to direct imaging with next-decade telescopes, such as the Wide-Field Infrared Space Telescope \citep[WFIRST;][]{Spergel2015arXiv} and the European Extremely Large Telescope \citep[ELT;][]{Quanz2015}. Another advantage of these systems being nearby is the larger orbital astrometric signature. While RV measurements allow us to estimate only the minimum mass of the planet ($m_p \sin i$), an astrometric orbit provides the inclination ($i$), and hence $m_p$.

Barnard's Star is the second-nearest stellar system. The search for planets around Barnard's Star depicts the history of achievable astrometric and RV precision for M dwarfs. It is also one of the most famous premature claims for exoplanet detections. For two decades \citet{vdK1963,vdK1969b,vdK1969a,vdK1975,vdK1982} claimed an astrometric detection of one or two Jovian planets with orbital periods between $11$ and $26$ years. These claims were first questioned by \citet{Gatewood1973} and later rejected at a confidence level of ${\sim}94\%$ by \citet{Choi2013}. By using precision RVs, \citet{Choi2013} were able to exclude the possibility of Jupiter-mass planets at almost any orbital period $\lesssim 25$ years, except for the most unlikely case of almost face-on orbits. The best existing astrometric constraints on planets around Barnard's Star were set by \citet{Benedict99}. Using the Fine Guidance Sensor of the Hubble Space Telescope (HST-FGS), \citet{Benedict99} managed to exclude an astrometric orbital perturbation larger than $1250$\,$\mu$as, with periods of $5$--$600$\,days, at ${\sim}95\%$ confidence level, which translates to mass upper limits of ${\sim}1$\,M$_J$ at $P\sim 150$ day orbit or ${\sim}0.5$\,M$_J$ at $P\sim 400$ day orbit.

Recently, \citet{Ribas2018} have reported an RV detection of a planet candidate in orbit around Barnard's Star. Table 1 gives the main parameters of Barnard's Star and the detected planet. In this paper we ask the question whether the true mass of the planet can be constrained by accurate astrometric measurements of Barnard's Star. In what follows, we estimate the probability of detecting the astrometric signature of the planet with existing and upcoming instruments, as well as the achievable mass upper limit in case of no detection.

\section{Current and upcoming astrometric capabilities}
\label{sec2}

When it comes to detecting planetary-induced astrometric orbits, with typical astrometric signatures of $<1000$\,$\mu$as, the most important figure of merit is the single-epoch astrometric accuracy achievable by an instrument ($\sigma_\Lambda$). Ground-based differential astrometry is generally limited by atmospheric turbulence \citep{Sahlmann2013SPIE}. Although for subarcsecond binaries, a $\sigma_\Lambda$ of ${\sim}10$\,$\mu$as can be achieved with optical interferometry \citep[e.g.,][]{Lane2004}, for single stars imaging astrometry can only achieve a $\sigma_\Lambda$ of ${\sim}100$\,$\mu$as with instruments such as the FORS2 camera on the Very Large Telescope \citep[VLT-FORS2;][]{Lazorenko2014,Sahlmann2014}. For Barnard's Star, however, performance might be degraded to $\sigma_\Lambda\sim300$\,$\mu$as because of the scarcity of bright enough reference stars in the FORS2 field of view \citep[e.g,][]{Sahlmann2016}. Next-decade $30$ m class telescopes with wide-field correction adaptive optics might bring these numbers closer to the theoretical atmospheric limits of ${\sim}40$\,$\mu$as \citep[e.g.,][]{Trippe2010}. However, astrometric characterization of bright stars, such as Barnard's Star, would most efficiently be done from space.

Hipparcos astrometric measurements have one-dimensional $\sigma_\Lambda$ of ${\sim}700$\,$\mu$as for the brightest stars and typically ${\sim} 1500$\,$\mu$as for $V=9$\,mag stars, such as Barnard's Star \citep{Perryman1997,vanLeeuwen2007,vanLeeuwen2007b}. Astrometric measurements with the HST-FGS provided $\sigma_\Lambda$ of $100$--$300$\,$\mu$as for many targets over the last two decades \citep{Benedict2017}. A new approach that uses the HST Wide Field Camera 3 (HST-WFC3) in spacial-scanning mode demonstrated $20$--$80$\,$\mu$as astrometric accuracy on bright stars \citep{Riess2014,Riess2018,Casertano2016}. Specifically for Barnard's Star, however, the low number of bright enough reference stars in the WFC3 field of view means that achieving the highest possible accuracy with this method heavily depends on the ability to predict the variations of the geometric distortion of the detector produced by the thermal cycle of HST \citep{Riess2014,Riess2018}. An optimal filter choice and an improved optical model of the telescope may eventually lead to $\sigma_\Lambda\sim30$\,$\mu$as.

Barnard's Star is currently being observed by Gaia, among another ${\sim} 1.3$ billion stars of our Galaxy \citep{Gaia2018}. By the end of $2018$ it should have transited Gaia's focal plane $>50$ times\footnote{https://gaia.esac.esa.int/gost/index.jsp}, and by the end of $2022$ the number of observations ($N_{\rm obs}$) will grow to $\gtrsim100$. However, about half of these scans are not individual-epoch observations, but rather adjacent transits of Gaia's two fields of view, separated by a couple of hours. In addition, since Gaia performs global astrometry, the individual measurements of the highest possible accuracy will not be available before the final data release \citep{Lindegren2016}. The expected along-scan uncertainty for the $G\simeq8.2$\,mag Barnard's Star varies from ${\sim}34$\,$\mu$as \citep{Perryman2014}, through ${\sim}50$\,$\mu$as \citep{Sozzetti2014}, to $\gtrsim 100$\,$\mu$as \citep{Lindegren2018}. The different estimates emerge from differences in the assumed levels of centroid, attitude, and calibration errors, and by the assumed impact of Gaia's gating scheme on bright-star astrometry.

For the next decade NASA is planning to launch WFIRST. \citet{WFIRST2017arXiv} estimated that it will enable $\sigma_\Lambda\sim10$\,$\mu$as astrometry, by using two different techniques: spatial scanning and diffraction spike modeling \citep[see also][]{Melchior2018}. We consider this estimate as rather optimistic.

For the purpose of comparing the different astrometric instruments as follow-up tools to detect the planetary-induced orbit of Barnard's Star, we assume for each instrument an optimistic--pessimistic range of the achievable single-epoch accuracy ($\sigma_\Lambda$). We take this range to be $100$--$300$\,$\mu$as for the VLT, $30$--$100$\,$\mu$as for Gaia and the HST, and $10$--$30$\,$\mu$as for WFIRST. Additional issues that should be taken into account when comparing the different astrometric instruments are discussed in Sect. \ref{sec4}.

\section{Astrometric signature of Barnard's Star b}
\label{sec3}

\begin{table}
\caption{Main parameters of Barnard's Star and Barnard's Star b.}
\begin{tabular}{lccc}
\hline
\hline
\noalign{\smallskip}
Parameter name & symbol & value & ref. \\
 & (units) & & \\
 \noalign{\smallskip}
\hline
\noalign{\smallskip}
Barnard's Star & & & \\
\noalign{\smallskip}
\hline
\noalign{\smallskip}
mass & $m_\star$ (M$_{\odot}$) & $0.16\pm0.02$ & $1,3$ \\
center-of-mass RV & $\gamma$ (km\,s$^{-1}$) & $-110.5\pm0.1$ & $2,3$\\
G magnitude & G (mag) & $8.195\pm0.002$ & $4$\\
parallax & $\varpi$ (mas) & $547.45\pm0.29$ & $4$ \\
proper motion in RA & $\mu_\alpha$ (mas/yr) & $-802.8\pm0.6$ & $4$ \\
proper motion in DEC & $\mu_\delta$ (mas/yr) & $10362.5\pm0.4$ & $4$ \\
\noalign{\smallskip}
\hline
\noalign{\smallskip}
Barnard's Star b & \\
\noalign{\smallskip}
\hline
\noalign{\smallskip}
orbital period & $P$ (day) & $232.8\pm0.4$ & $1$ \\
RV semiamplitude & $K$ (m\,s$^{-1}$) & $1.20\pm0.12$ & $1$ \\
eccentricity & $e$ & $<0.42$ & $1$ \\
argument of periastron & $\omega$ (deg) & $106\pm21$ & $1$ \\
\noalign{\smallskip}
min. mass & $m_p\sin i$ (M$_\oplus$) & $3.23\pm0.44$ & $1$ \\
min. astrometric signature & $\alpha\sin i$ ($\mu$as) & $13.3\pm1.3$ & $1$ \\
\noalign{\smallskip}
\hline
\end{tabular}
\tablebib{(1) \citet{Ribas2018} (2) \citet{Nidever2002}; (3) \citet{Reiners2018b}; (4) \citet{GaiaDR2};
}
\label{tab1}
\end{table}

The astrometric motion of a planet-hosting star is composed of three major components: parallactic, proper, and orbital. The amplitude of the latter is usually referred to as the astrometric signature of the planet and is given by
\begin{equation}
\alpha = \bigg(\frac{a_{\rm app}}{\rm 1 AU}\bigg) \cdot \varpi,
\label{eq1}
\end{equation}
where $a_{\rm app}$ is the semimajor axis of the apparent orbit of the star, and $\varpi$ is the parallax \citep[e.g.,][]{Reffert2011}.

For RV-detected planets, a minimum barycentric semimajor axis of the host star, $a_\star \sin i$, and the planet's minimum mass, $m_p \sin i$, can be derived from the orbital parameters via\footnote{The term `minimum' refers to the unknown inclination.}
\begin{equation}
2\pi a_\star \sin i = KP(1-e^2)^{1/2}, {\rm and}
\label{eq2}
\end{equation}
\begin{equation}
2\pi G m_p \sin i \simeq KP^{1/3}(1-e^2)^{1/2}m_\star^{2/3},
\label{eq3}
\end{equation}
where $G$ is the universal gravitational constant, and the other parameters are defined in Table \ref{tab1} \citep[e.g.,][]{Eggenberger2010}. The approximation in Eq. \ref{eq3} is valid for $m_p \ll m_\star$.

For a star at a known distance, assuming an isotropic distribution of orbital orientations, which induces a known probability distribution of the inclination, we can estimate the probability of the astrometric signature to be in a certain range of values. In particular, we can estimate the probability of the astrometric signature to be above the detection thresholds of the various available instruments. Moreover, owing to the close relation between $\sin i$ and the expected astrometric signature of a known RV planet, a nondetection at a given threshold can be translated to an inclination lower limit, which can be directly translated to an upper limit on the mass of the planet if the mass of the star is known.

By combining numerous measurements from precision RV instruments, \citet{Ribas2018} have revealed a low-amplitude periodic signal in the RVs of Barnard's Star, which is best explained by a planetary companion with a minimum mass of ${\sim}3.2$\,M$_\oplus$ at an orbital period of ${\sim}233$ days. The orbital-parameter uncertainties, listed in Table \ref{tab1}, were estimated by using Markov Chain Monte Carlo (MCMC) analysis. In order to estimate the detectability of Barnard's Star astrometric signature, we used the RV-analysis MCMC chain of \citet{Ribas2018}. For each point in the chain we calculated $a_\star\sin i$ using Eq. \ref{eq2}. In addition, for each MCMC point we drew an inclination from an isotropic orbital-orientation distribution, i.e., from a $\sin i$ probability density function (PDF). By doing so, we actually assumed a flat prior for the mass of Barnard's Star b \citep[e.g.,][]{HoTurner2011,LopezJenkins2012}. Given the as yet unknown mass distribution of planets in orbital periods of $>200$\,days around M dwarfs \citep[e.g.,][]{DressingCharbonneau2015}, we believe this is a reasonable assumption.

\begin{figure}
\minipage{0.5\textwidth}
{\includegraphics[width=\linewidth]{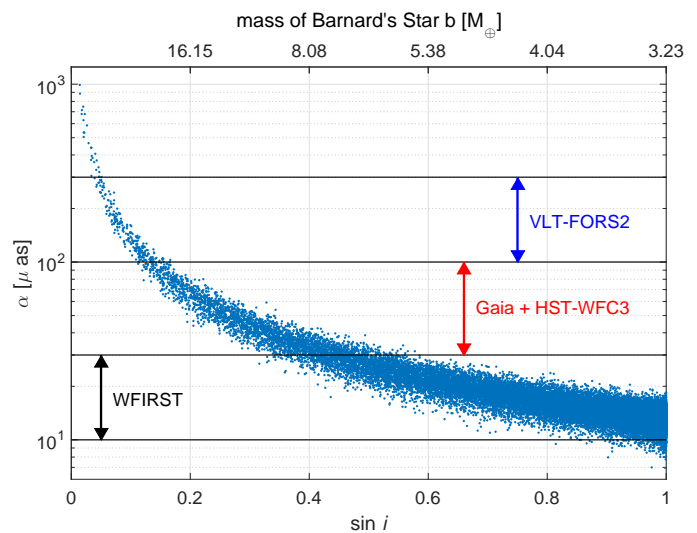}}
\endminipage\hfill
\caption{Astrometric signature of Barnard's Star b, calculated from the RV-analysis MCMC chain of \citet{Ribas2018}, as a function of the simulated $\sin i$ chain. The horizontal lines indicate the detection thresholds discussed in the main text. The pessimistic--optimistic single-epoch accuracy range of the four instruments discussed in the main text are specified with the colored arrows. The plot is truncated at $1250$\,$\mu$as following the upper limit set by \citet{Benedict99}. As a reference, the planet mass corresponding to each $\sin i$ value (assuming $m_p\sin i = 3.23$\,M$_\oplus$) is indicated at the top of the panel.
}
\label{fig1}
\end{figure}

For each MCMC+inclination chain point, we translated the orbital parameters to the semimajor axis of the apparent orbit ($a_{\rm app}$) by using the prescription given in \citet{Reffert2011}, App. A. The calculation uses $a_\star, \omega, i, e,$ and $\Omega$, the latter being the longitude of the ascending node. Since $\Omega$ cannot be estimated from the RVs, but the resulting $a_{\rm app}$ is independent of its value, we arbitrarily chose $\Omega=\pi$. We then translated $a_{\rm app}$ to $\alpha$ by using Eq. \ref{eq1}. We thus converted the RV-analysis results to a PDF of $\alpha$.

\begin{table}
\caption{Detection probabilities for the astrometric signature of Barnard's Star b, and mass upper limits in case of no detection, for four different detection thresholds, and their association with either the optimistic or the pessimistic single-epoch accuracy of the four astrometric instruments discussed in the main text.}
\begin{tabular}{cccl}
\hline
\hline
$T_{\rm det}$ & $P_{\rm det}$ & $m_{\rm up}$ & Instrument \\
($\mu$as)     &        ($\%$)      &  (M$_\oplus$)  & (-opt = optimistic; -pes = pessimistic)\\
\hline
$300$ & $0.1$  & $72.6$ & VLT-FORS2-pes \\
$100$ & $1.0$  & $24.3$ & Gaia-pes, HST-pes, VLT-FORS2-opt \\
$30$ & $10.5$  & $7.9$ & Gaia-opt, HST-opt, WFIRST-pes \\
$10$ & $99.1$ &  $3.2$ & WFIRST-opt \\
 \hline
\end{tabular}
\label{tab2}
\end{table}

In order to illustrate the process, Fig. \ref{fig1} shows the resulting chain of $\alpha$ values as a function of the simulated $\sin i$ chain. We then used the $\alpha$ PDF to estimate the probability of detecting the astrometric signature of the planet ($P_{\rm det}$) as the fraction of chain points with $\alpha$ above a certain detection threshold ($T_{\rm det}$). The horizontal lines in Fig. \ref{fig1} represent the selected detection thresholds, which can be associated with either the optimistic or the pessimistic $\sigma_\Lambda$ of four of the different astrometric instruments discussed in Sect. \ref{sec2}. In Table \ref{tab2} we give the estimated $P_{\rm det}$ for the four possible $T_{\rm det}$ values, and their association with the different astrometric instruments.

For the case of no detection we estimated the mass upper limit ($m_{\rm up}$) in a similar fashion. For each MCMC+inclination chain point we derived the corresponding planet mass using Eq. \ref{eq3}, thus creating a PDF of $m_{\rm p}$. Each $T_{\rm det}$ separates the $m_{\rm p}$ PDF in two, slightly overlapping, PDFs: one for detectable orbits and one for nondetectable orbits. We then estimated $m_{\rm up}$ as the highest possible mass for the nondetectable orbits. The values of $m_{\rm up}$ for the different values of $T_{\rm det}$ are also given in Table \ref{tab2}.

\section{Discussion}
\label{sec4}

The two existing most promising instruments to detect the astrometric signature of Barnard's Star b are Gaia and the HST. However, both have intrinsic systematic issues that might prevent them from achieving their best performance for this task. Moreover, Barnard's Star has specific properties that complicate the astrometric confirmation. We now address the issues that would have the largest impact on the astrometric-signature detectability and discuss possible mitigation strategies.

In Sect. \ref{sec3} we assigned $T_{\rm det}=\sigma_\Lambda$. This assignment assumes that an astrometric orbit can be detected for $\alpha \gtrsim \sigma_\Lambda$. Considering the case of unknown orbital periods, \citet{Casertano2008} required $\alpha \gtrsim 3\sigma_\Lambda$ to detect a planetary orbit. For RV-detected planets, however, \citet{Sahlmann2011} showed that the astrometric signature can be detected with Hipparcos data if $\alpha \gtrsim \sigma_\Lambda$. More specifically, they showed that the correct orbital parameters are derived when the orbit is detected at a significance of $>3\sigma$, which can be achieved for an astrometric signal-to-noise ratio of $S/N = \alpha\cdot\sqrt[]{N_{\rm obs}}/\sigma_\Lambda > 7$ \citep[see also][]{Sahlmann2016}. Assuming a similar relation for both Gaia and HST, we conclude that an $\alpha=\sigma_\Lambda$ planetary-induced orbits of known RV planets will be detected with $N_{\rm obs}\gtrsim 50$. For Gaia, $N_{\rm obs}$ is defined by the mission duration and scanning law. The mission extension to end of $2022$ will indeed bring the number of individual-epoch observations of Barnard's Star to $\gtrsim50$. In HST there is the advantage that $N_{\rm obs}$ can be tuned to achieve a desired detectability once the actual $\sigma_\Lambda$ has been determined from preliminary observations \citep[e.g.,][]{Riess2014}.

Another important aspect of $N_{\rm obs}$ and of observations scheduling is the need to simultaneously constrain several parameters, some of which are correlated among each other. For a single planet around Barnard's Star, the astrometric model is a nonlinear function of $14$ parameters: the seven standard astrometric parameters (including secular acceleration\footnote{We assume the change in secular acceleration will be negligible in the timespan of HST and Gaia observations.}), and another seven orbital parameters for the planet \citep[e.g.,][]{Sahlmann2011}. Five of the seven orbital parameters are common to astrometry and RV, so we are left with nine parameters to be determined only by the astrometric measurements: $\alpha,\delta,\mu_\alpha,\mu_\delta,\varpi,\dot{\mu}_\alpha,\dot{\mu}_\delta,\Omega,$ and $i$. The annual change in parallax \citep[$\dot{\varpi}\sim34$\,$\mu$as\,yr$^{-1}$;][]{Dravins1999} and the secular acceleration due to change in perspective \citep[$\dot{\mu}\sim 1.2$\,mas\,yr$^{-2}$;][]{Benedict99} are not independent parameters, but rather functions of $\varpi,\mu,$ and the absolute RV. Therefore, $N_{\rm obs}\sim50$ should give enough degrees of freedom to constrain all nine parameters.

Barnard's Star's high proper motion may degrade the achievable precision in narrow-field relative astrometry. Over time, it might lead to changing the set of reference stars. This is a major problem for VLT and HST astrometry, but not for Gaia. In addition, Barnard's Star's ${\sim}1\%$ photometric variability \citep{Benedict1998} might induce astrometric jitter. However, it is not expected to be larger than a few $\mu$as \citep[e.g.,][]{Eriksson2007}.

\citet{Ribas2018} have also mentioned that a particular way of combining the measured RVs leads to the detection of an additional modulation with a period $>10$\,years. Although this long-term perturbation most likely arises from a magnetic activity cycle, its interpretation as an additional planet with $m_p\sin i\sim15$\,M$_\oplus$ is not ruled out. The minimum astrometric signature of such a massive outer planet would be ${\sim}500$\,$\mu$as, which can be detected by any of the instruments mentioned in Table \ref{tab2}.

\section{Summary and conclusions}
\label{sec5}

We reviewed the astrometric capabilities and limitations of current and forthcoming instruments, and found Gaia and the HST to be currently the most promising instruments to detect the astrometric signature of Barnard's Star b. Taking the optimistic estimate of their single-epoch accuracy of $\sim30$\,$\mu$as, we found a probability of ${\sim}10\%$ to detect the planet's astrometric signature with $N_{\rm obs}\sim50$. In case of no detection, which would correspond to a nearly edge-on orbit, the implied mass upper limit would be $m_{\rm p}\lesssim8$\,M$_\oplus$, which would place the planet in the super-Earth mass range.

Despite the fact that Gaia will continue observing Barnard's Star for the next few years and will release its results around $2023$, we expect additional astrometric HST-WFC3 follow-up observations to improve the constraints on the mass of Barnard's Star b. The timings and scan directions of HST observations can be tuned to help determining the free parameters of the full astrometric model. If in the next decade an accuracy of $\sigma_\Lambda\sim 10$\,$\mu$as is indeed reached for Barnard's Star, for example with WFIRST \citep{Melchior2018}, the prospects of measuring the true mass of the planet will grow to ${\sim}99\%$. Then, Gaia and HST observations performed in the next few years will set valuable constraints on some parameters of the astrometric model that benefit from observations over a long time baseline.

In the coming few years RV surveys of nearby stars, such as CARMENES \citep{Quirrenbach2018} and Reddots\footnote{https://reddots.space/}, are expected to detect dozens of Earth-to-Neptune mass planets with orbital periods $\lesssim 1$\,year. These planets will be excellent targets for characterization with existing and upcoming complementary techniques, such as direct imaging and astrometry.


\begin{acknowledgements}
This research was supported by the ISRAEL SCIENCE FOUNDATION (grant No. 848/16), with additional contributions by the DFG Research Unit FOR2544 “Blue Planets around Red Stars”. G.A-E research is funded via the STFC Consolidated Grant ST/P000592/1, and a Perren foundation grant.
\end{acknowledgements}

\bibliographystyle{aa}
\bibliography{LevTalOr.bib}


\end{document}